\documentclass[11pt, reqno]{amsart}
\usepackage{amsmath}
\usepackage{amsthm}
\usepackage{mathrsfs}
\usepackage{amssymb}
\usepackage{graphicx}
\usepackage[round]{natbib}

\usepackage{multirow}
\usepackage{array}
\usepackage{booktabs}
\usepackage{lmodern}
\usepackage[colorlinks, citecolor=blue, linkcolor=blue, urlcolor=blue]{hyperref}
\usepackage{enumitem}
\usepackage{subcaption}

\theoremstyle{plain}

\theoremstyle{definition}

\theoremstyle{remark}

\newcommand{\bvarepsilon}{\mbox{\boldmath$\varepsilon$\unboldmath}}

\newcommand{\bbeta}{\mbox{\boldmath$\beta$\unboldmath}}

\title[
Analysis of Gaussian Spatial Models with ...]{
Analysis of Gaussian Spatial Models with Covariate Measurement Error}
\author[Tadayon, V.]{Vahid Tadayon\\ Department of Statistics, Eghlid Higher Education Institute, Eghid, Iran. \\ Email: vahidtadayon24@gmail.com %$^{\dagger}$
}
\begin{document}
\bibliographystyle{plainnat}
\maketitle
%\footnotetext[1]{\ Corresponding author}

\begin{abstract}
Uncertainty is an inherent characteristic of biological and geospatial data which is almost made by measurement error in the observed values of the quantity of interest. Ignoring measurement error can lead to biased estimates and inflated variances and so an inappropriate inference. In this paper, the Gaussian spatial model is fitted based on covariate measurement error. For this purpose, we adopt the Bayesian approach and utilize the Markov chain Monte Carlo algorithms and data augmentations to carry out calculations. The methodology is illustrated using simulated data.
\end{abstract}

\noindent 

\section{Introduction}\label{s1}
Particulate matter (PM, hereafter), which includes the harmful suspended mixture of both solid and liquid particles, is generally encompassed in air pollution. They are often separated into three classifications: ``coarse", ``fine" and ``ultrafine" particles. Coarse particles have a diameter of between $10$ micrometer ($\mu m$) and $2.5\mu m$ and settle relatively quickly whereas fine ($0.1$ to $2.5\mu m$ in diameter) and ultrafine ($<0.1\mu m$ in diameter) particles remain in suspension for longer. To put things into perspective, human hair has a diameter of $50-70\mu m$ and a grain of sand has a diameter of $90\mu m$.

When someone talks about ${\rm PM}_{10}$, they are referring to particles smaller than $10\mu m$. ${\rm PM}_{10}$ is a mixture of materials that can include smoke, soot, dust, salt, acids, and metals. There are sources of ${\rm PM}_{10}$ in both urban and rural such as: Motor vehicles, wood burning stoves and fireplaces, dust from construction, landfills, and agriculture, wildfires and brush/waste burning, and industrial sources. ${\rm PM}_{10}$ is among the most harmful of all air pollutants. When inhaled these particles evade the respiratory system's natural defenses and lodge deep in the lungs. ${\rm PM}_{10}$ can increase the number and severity of asthma attacks, cause or aggravate bronchitis and other lung diseases and reduce the body's ability to fight infections. Moreover, skin as a largest organ in body, acts as the first and most important defense barrier against environmental contaminants. Skin has numerous pores, which are decidedly larger than PM, there is no direct evidence that PM can penetrate into skin regardless of smaller size. However, it is undoubtedly reported that the particles can penetrate skin though hair follicles depending on their size, indicating the penetration of PM through transfollicular route (for more information, see  \cite{Lademann, Davidson, Harrison1, Harrison2, Wang}).

Since an excess of ${\rm PM}_{10}$ concentration in any area causes serious environmental pollution, the detection of areas with high ${\rm PM}_{10}$ concentration is an important problem and this is a greatest concern of recent studies that link ${\rm PM}_{10}$ exposure to the premature death of people who already have heart and lung disease, especially the elderly. Due to ${\rm PM}_{10}$ concentrations as a geostatistical data are spatially correlated such that observations closer in space tend to be more correlated than observations farther away in space, a statistical model which incorporates spatial information have to be employed for detecting areas with excess of ${\rm PM}_{10}$ concentrations in air pollution.

In spatial data modeling, it is commonly assumed that the covariates have been observed exactly. When this assumption is violated due to the measurement technique or instruments used, the results can raise interpretation issues. However, some recently published work has suggested that to accommodate measurement error in a spatial context. Some of these works include  \cite{Gryparis},  \cite{Sheppard}, \cite{tadayon2015bayesian}, \cite{tadayon2017bayesian}, \cite{tadayon2018NonGaussian} and  \cite{tadayon2018Spatial}. In most of the aforementioned works which the inference was conducted based on a frequentist approach, concerns have been raised regarding the computational complexity of the proposed algorithms for estimating model parameters. Moreover, the maximum likelihood estimators are associated with larger variances \citep{Li}. 

In this article, we aim to fit a linear model on the spatial data with covariate measurement errors. Since the likelihood function is complex, the Bayesian approach is adopted for statistical inference. However, the proposed model provides flexibility in capturing, it facilitates representing and taking fuller account of the uncertainties related to models and parameter values and we able to incorporate prior information. Furthermore, the Bayesian approach, via the use of a weakly informative prior, also provides estimates with good frequentist coverage properties and provides us with a correct variance estimate \citep{Gryparis2}. On the other hand, the main advantage of the Bayesian approach consists in providing the full posterior distribution, which provides a powerful tool for inference \citep{Berry}. Therefore, the Bayesian approach is much more computationally intensive than the others are. 

The organization of the paper is as follows. After describing our proposed model and the notations (Section \ref{s2}), in continuing, we describe the details of the Bayesian inference and prediction. An example with simulated data is presented in Section \ref{s3}. Section \ref{s4} we illustrates implementation details in a fully Bayesian framework and the results on modeling of covariate measurement error in ${\rm PM}_{10}$ concentrations. The article ends with a brief conclusion section.

\section{Model formulation and Inference}\label{s2}
Consider modelling a phenomenon of interest as a random process $Y\left(s\right)$ at location $s$ in the spatial region. Assume that we observe a realization of this process such as ${\bf y}= \left( {{y_1},{y_2}, \ldots ,{y_n}} \right)'$ at locations $s_i$ for $i =1,2,\ldots,n$ and write the model for the $i$th location given the covariates ${\bf x}$ as
\begin{eqnarray}\label{E1}
Y\left( {s_i} \right) = {\beta _0} + {\bf x}'{\left( {s_i} \right)}{\bbeta_x}   + {\sigma  }\varepsilon \left( s \right) + {\sigma _\rho }\rho \left( s \right),
\end{eqnarray}
where the mean surface ${\bf x}'{\left( \cdot \right)}{\bbeta_x}$ with ${\bbeta_x} = \left( {{\beta_1},{\beta_2}, \ldots ,{\beta_k}} \right)'$ is often termed \textit{trend} or \textit{drift}.  $\varepsilon \left(\cdot\right)$ is a zero-mean Gaussian random field with one variance and isotropic correlation function
\begin{eqnarray*}
Corr\left\{ {\varepsilon \left( {{s_i}} \right),\varepsilon \left( {{s_j}} \right)} \right\} = {C_\theta }\left( {\left\| {{s_i} - {s_j}} \right\|} \right) = {C_\theta }\left( {\left\| h \right\|} \right)
\end{eqnarray*}
that depends on an unknown $q\times 1$ parameter $\theta$ and independent of the uncorrelated Gaussian white noise process $\rho \left(\cdot\right)$ with zero-mean and variance 1. The covariance matrix $C_\theta\left(\cdot\right)$ admits many choices. A widely adopted choice for this correlation function is the Matérn function which includes the exponential model and the Gaussian correlation model as two special cases \citep{Cressie}. In the presence of measurement error the covariate ${ X}\left(\cdot\right)$ cannot be observed directly, but ${ W}\left(\cdot\right)$ that are surrogates for the ${X}\left(\cdot\right)$ are observed and we can write ${ W}\left(\cdot\right)=X\left(\cdot\right)+{ U}\left(\cdot\right)$, with $E\left[{{ U}\left(\cdot\right)}\right]=0$. In this setting, a distinction is made between the ``functional" case where the elements of $X$ are treated as fixed values and the ``structural" case where they are random. Here unlike Li {\it et al.} \cite{Li}, we consider the functional case and so we have
\begin{eqnarray}\label{E2}
Y\left( s \right) = { W}\left( s \right)\bbeta  + {\sigma  }\varepsilon \left( s \right) + {\sigma _\rho }\rho \left( s \right) - { U}\left( s \right)\bbeta .
\end{eqnarray}
In this case, the existence of the correlation between ${ W}\left(\cdot\right)$ and ${ U}\left(\cdot\right)$ is problematic and affects the analysis. To overcome this obstacle, following \cite{Carroll}, we provide the alternative model
\begin{eqnarray}\label{E3}
Y\left( s \right) &=& \underbrace {E\left[ {X\left( s \right)\left| {W\left( s \right)} \right.} \right]}_{\mu \left( s \right)} \bbeta  + {\sigma  }\varepsilon \left( s \right) + {\sigma _\rho }\rho \left( s \right) + {\sigma _v } V\left( s \right)\bbeta \nonumber\\
&=&\mu \left( s \right)\bbeta  + {\sigma  }\varepsilon \left( s \right) + {\sigma _\rho }\rho \left( s \right) + {\sigma _v } V\left( s \right)\bbeta,
\end{eqnarray}
in which by considering the fact that $X$ is fix, we substituted $E\left[ {X\left(\cdot\right)\left| {W\left(  \cdot  \right)} \right.} \right] + V\left(  \cdot  \right)$ for  $W\left(\cdot\right)$ in (\ref{E2}), where, $V\left(\cdot\right)=E\left[{U\left(\cdot\right)\left| W\left(\cdot\right) \right.} \right]-U\left(\cdot\right)$ and clearly, $E\left[ {V\left(  \cdot  \right)} \right] = 0$. It should be mentioned that in the proposed model (\ref{E3}) three random fields $\varepsilon \left(\cdot\right)$, $\rho \left(\cdot\right)$ and $V\left(\cdot\right)$ are considered independent of each other. The scale parameters $\sigma$, $\sigma_\rho$ and $\sigma_v$ all three are defined in ${\Re ^+}$ and the ratio ${\omega ^2} = \frac{{\sigma _\rho ^2}}{{\sigma  ^2}}$ indicates the relative importance of the nugget effect. If ${\tau ^2} = \frac{{\sigma _v^2}}{{{\sigma ^2}}}$ and $\eta ' = \left( {\bbeta ,\sigma  ^2,{\omega ^2} ,{\tau^2},\theta } \right)$ represents the vector of model parameters, the joint distribution of $Y$ given the vector $\eta$ can be written as
\begin{eqnarray*}
\underline{Y}\left(  s  \right) \left| \eta\right.\sim {N_n}\left( {\mu \left( s \right)\bbeta ,{\sigma ^2}\left[ {{C_\theta } + \left( {{\omega ^2} + {\tau ^2}{\bbeta}'\bbeta} \right){I_n}} \right]} \right).
\end{eqnarray*}
%and the likelihood function by marginalizing is given by
%\begin{eqnarray}
%L\left( {\eta \left| \underline{y} \right.} \right) = \int_{{{\Re}^{\hspace{-0.7mm}\frac{}{}p}}} {P\left({\underline{Y}\left|{{\bf v},\eta}\right.}\right)P\left({{\bf v}\left|\eta\right.}\right)d{\bf v}},
%\end{eqnarray}
%in which the objective of ${\bf v}=\left(v_1,v_2,\ldots,v_p\right)$ is the vectorized version of the matrix $V$.
%

We will now outline the Bayesian analysis of the proposed spatial model. To specify the model from a Bayesian point of view, we need to select the prior distributions for all unknown parameters. In the absence of prior information, a convenient strategy of avoiding improper posterior distribution is to utilize proper (but diffuse) priors. However, in this case, the prior hyperparameters may have an unpleasant influence on inference and so; the insignificance of the prediction sensitivity to the hyperparameters should be established. In the Bayesian analysis of spatial models, it is usually assumed that, the model parameters are mutually independent, thus for convenience but not necessary optimal, here we apply this assumption, $\pi \left( {\bbeta ,\sigma  ^2,{\omega ^2} ,{\tau^2},\theta } \right) = \pi \left( \bbeta  \right)\pi \left( {{\sigma ^2}} \right)\pi \left( {{\omega ^2}} \right)\pi \left( {\tau^2}  \right)\pi \left( \theta  \right)$. In what follows, we will describe the adopted prior distributions and list some alternatives. The hyperparameters of the adopted priors including ${c_1},{c_2},\ldots,{c_9}$ and $\gamma$ which are chosen to reflect vague prior information.

{\it Prior on $\bbeta$.} A conjugate prior for $\bbeta$ is the multivariate normal distribution ${N_p}\left( {{\bf 0},{c_1}{I_p}} \right)$, which implies that the posterior is multivariate normal as well. Here are some alternative choices: matrix-normal density and  ridge regression prior \citep{Minka}, power prior \citep{Ibrahim} and normal-gamma prior \citep{Griffin}.

{\it Prior on $\sigma^{-2}$.}  Ideally, for variance parameter, we often expect a prior that is invariant to rescaling the observations. Because that cannot be achieved exactly with a proper prior, we approximate this by adopting $\sigma ^{-2} \sim Gamma\left( {{c_2},{c_3}} \right)$ with very small values of $c_2$ and $c_3$. The use of half-normal prior for variance parameters also has been studied by \cite{Gelman} and \cite{Lambert}

{\it Prior on $\omega^{2}$.} For $\omega^{2}$, we often expect values that are smaller than unity, and most prior mass could be around small values (no nugget effect), so that any strong nugget effect comes from the data rather than the prior. However, we do not want the prior to exclude a large nugget effect either. Thus, we propose using the flexible generalized inverse-Gaussian (GIG) prior $GIG\left( {\gamma ,{c_4},{c_5}} \right)$ with density function
\begin{eqnarray*}
f\left( {{\omega ^2};\gamma ,{c_4},{c_5}} \right) = \frac{{{{\left( {\frac{{{c_5}}}{{{c_4}}}} \right)}^\gamma }{\left( {{\omega ^2}} \right)^{\gamma  - 1}}}}{{2{\kappa _\gamma }\left( {{c_4}{c_5}} \right)}}{e^{ - \frac{1}{2}\left\{ {c_4^2{{\left( {{\omega ^2}} \right)}^{ - 1}} + c_5^2{\omega ^2}} \right\}}},\quad {c_4},{c_5} \in {{{\Re}}^ + }, \gamma  \in \Re 
\end{eqnarray*}
where $\kappa _\gamma$ is the modified Bessel function of the third kind \citep{Barndorff}. Further details on GIG distribution and the special cases can be found in \cite{Bibby}. 

{\it Prior on $\tau^{2}$.} By the same idea expressed in $\omega^{2}$, we let $\tau^2  \sim GIG\left( {0,{c_6},{c_7}} \right)$.

{\it Prior on $\theta$.} The Matérn correlation function is parameterized in terms of $\theta =\left( {{\theta _1},{\theta _2}} \right)$. The prior on ${\theta _1}$ should take into account that the value of this range parameter is critically dependent on the scaling of the distance $d$. In fact, the correlation structure depends only on ${\theta _1}$ through ${{{\theta _1}} \mathord{\left/ {\vphantom {{{\theta _1}} d}} \right. \kern-\nulldelimiterspace} d}$. We propose ${\theta _1}\sim Exp\left( {{{{c_8}} \mathord{\left/ {\vphantom {{{c_8}} {med\left( d \right)}}} \right.\kern-\nulldelimiterspace} {med\left( d \right)}}} \right)$, where ${med\left( d \right)}$ is the median value of all distances in the data. The smoothness parameter ${\theta _2}$ is linked to the degree of mean square differentiability of the process and will be given as ${\theta _2} \sim Exp\left( {{c_9}} \right)$.

Now, combining the likelihood function $L\left( {\eta \left| {\bf{y}} \right.} \right)$ and the prior distribution $\pi\left({\bbeta,\sigma^2,{\tau^2},\nu,\theta}\right)$, the posterior distribution of parameters can be obtained. To facilitate the sampling, we first augment with the latent variables $V$ and the vector $\varepsilon$ defined in (\ref{E1}). Thus we try to draw samples from all unknown quantities ${\footnotesize P\left( {\eta,\bvarepsilon , V\left|\underline{y} \right.} \right)}$ via MCMC methods such as Gibbs sampler and Metropolis-Hastings algorithm. Details of the MCMC algorithm are presented in the Appendix. Hence, we produce some samples ${\footnotesize \left\{ {\eta^{\left( i \right)},{\bvarepsilon ^{\left( i \right)}},{V^{\left( i \right)}}} \right\}_{i = 1}^l}$ to implement the Bayesian inference. 

Since producing a map of contaminated areas requires predicting values at new locations, in sequel, we address to spatial prediction of the response variable $y_0=y\left(s_0\right)$ in unobserved location $s_0$ based on the Bayesian predictive distribution:
\begin{eqnarray}\label{E4}
P\left( {{Y_0}\left| \underline{y} \right.} \right)&=&  {\int_{{\Re^{n + 1}}} {\int_\Omega {P\left( {{Y_0},{V^ * },\eta\left| \underline{y} \right.} \right)} } } d\eta d{V^ * }\nonumber\\
&=& {\int_{{\Re^{n + 1}}} {\int_\Omega  {P\left( {{Y_0}\left| {{V^*},\eta ,\underline{y}}\right.} \right) } } } P\left( {{V_0}\left| {V,\eta ,\underline{y}} \right.} \right) P\left(V,\eta \left| \underline{y}\right. \right)d\eta d{V^*},\nonumber\\
\end{eqnarray}
where $\Omega$ denotes the parameter space of $\eta$ and $V_{{\left(n+1\right)}\times p}^*$ is the same as $V_{{n}\times p}$ but the vector ${{\underline{V_0}}}'$ as a row vector is incorporated before the first row of $V_{{n}\times p}$. Since
\begin{eqnarray*}
{\left( {{Y_0},\underline{Y}'} \right)^\prime }\left| {V^*},\eta  \right. \sim {N_{n+1}}\left( {\left[ \begin{array}{l}
\mu '\left( {{s_0}} \right)\\
~\mu \left( s \right)
\end{array} \right]\bbeta +{\sigma_v}{V^*}\bbeta
,{\sigma ^2}\left[ {C_\theta ^* +  {\omega ^2} {I_{n + 1}}} \right]} \right),
\end{eqnarray*}
where ${C_\theta ^*}=\left( {\begin{array}{*{20}{c}} 1&{{r_\theta }^\prime }\\ {{r_\theta }}&{{C_\theta }} \end{array}} \right)$ and for $i=1,2,\ldots,n$, ${{\left\lbrace {r_{{\theta}}}^\prime\right\rbrace }_i}  = {C_\theta }\left( {\left\| {{s_0} - {s_i}} \right\|} \right)$, then ${Y_0}\left|{V^*},\eta,\underline{y}  \right. $ has a univariate normal distribution with
\begin{eqnarray*}
E\left( {Y_0}\left|{V^*},\eta,\underline{y}  \right. \right) &=&\mu '\left( {{s_0}} \right)\bbeta  + \sigma \tau {v}'\left( {s_0}\right) \bbeta \\
&+&{{r_\theta }^\prime }C_\theta ^{ - 1}\left( {\underline{y}\left( {s}\right) -\mu \left( s \right)\bbeta  - \sigma \tau V\left( s \right)\bbeta } \right),\\
Var\left( {Y_0}\left|{V^*},\eta,\underline{y}  \right. \right) &=&{\sigma ^2}\left(1+ {{\omega ^2} } \right) - {r_\theta }^\prime C_\theta ^{ - 1}{r_\theta }.
\end{eqnarray*}
We also have ${V_0}\left| {\bvarepsilon ,V,\eta ,\underline{y} } \right. \sim  {N_p}\left( {{\bf 0},{I_p}} \right)$. Clearly, it can be also seen that sampling from the predictive distribution (\ref{E4}) is now straightforward: we directly draw the sample ${\footnotesize \left\{ {\eta^{\left( i \right)},{\bvarepsilon ^{\left( i \right)}},{V^{\left( i \right)}}} \right\}_{i = {l+1}}^{m_l}}$ from the posterior distribution described in the previous section after the burn-in point $l$, then we generate a drawing from conditional distribution $\footnotesize {V_0}\left| {\bvarepsilon ,V,\eta ,\underline{y} } \right.$ and finally using this sample, we can obtain a realization from the predictive distribution $\footnotesize P\left( {{Y_0}\left| \underline{y} \right.} \right)$. Repeating aforementioned steps as many times as required, thereby we generate a sample from the predictive distribution as $\left\{{{y_{{0_i}}}} \right\}_{i = l'+1}^{{m_{l'}}}$. Then the spatial predictor simply is given by averaging.

\section{Simulation study}\label{s3}
\begin{figure}
\centering
\includegraphics[width=0.75\textwidth, height=0.8\textwidth]{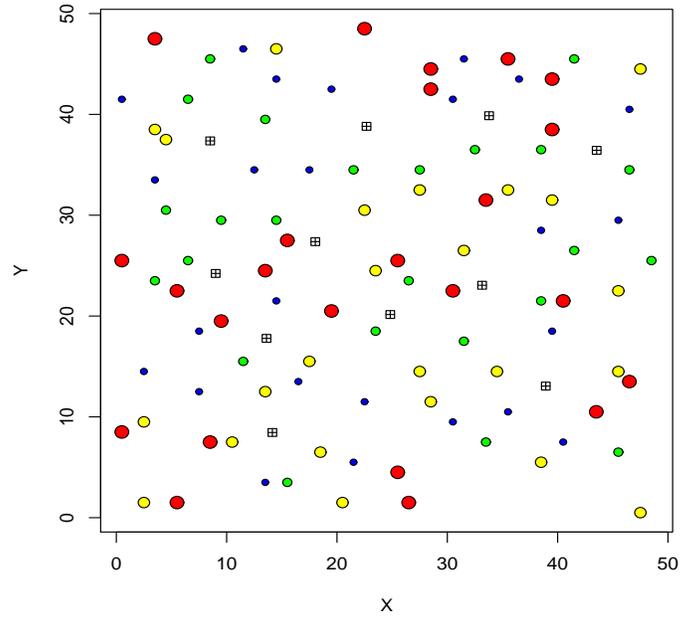}
\caption{Study region with the sampling locations and the generated data. A ``$\boxplus$" indicates the  hold-out dataset location.}
\label{f2}
\end{figure}

To examine the performance of our model, we conducted a simulation studies. For simplicity, we use coordinates of the data file $97data.dat$ available from GSLIB software \citep{Deutsch} in which 97 locations are taken on a pseudo-regular grid over a bidimensional region 50 by 50 miles. For model validation, we also selected 11 further locations as hold-out dataset and leave a sample size of n = 97 from which to fit the model. From Table \ref{t1}, one may observe the coordinates of these locations.
\begin{center}
[Table \ref{t1} about here.]
\end{center}
Our aims here are fourfold:
\begin{enumerate}[label=(\alph*)]
\item to evaluate MCMC performance and ensure that we obtained reasonable parameter estimates under the true model,
\item to assess the predictive performance of our methodology,
\item to illustrate the sensitivity of the proposed model to the benchmark values selected for hyperparameters and the basic values have been chosen for running the program
\item and the last aim is to examine identifiability of the parameters.
\end{enumerate}
First of all, we draw a sample of size $108$ from the normal distribution with mean $3$ and variance $0.2$ and save these values into a vector named $\bf{x}$. From now on, we will look at these data as a fixed (not random) and true values of a single covariate $x$. Then, we set parameters ${\beta_0}=0.5$, ${\beta_1}=2$, ${\sigma^2}= 1$, ${\omega ^2} = 1.1$ and using an isotropic exponential covariance function for $\theta=1.2$, we simulate a dataset on 108 considered locations based on the model 
\begin{eqnarray}\label{E5}
Y\left( {{s_i}} \right) = {\beta _0} + {\beta _1}{x \left( {{s_i}} \right)} + \sigma \varepsilon \left( {{s_i}} \right) + \sigma \omega {\rho \left( {{s_i}} \right)}.
\end{eqnarray}
Figure \ref{f2} shows a schematic description of the region that displays the sampling locations and the simulated dataset.
\begin{center}
[Figure \ref{f2} about here.]
\end{center}
In order to incorporate measurement error in the covariate $x$, after leaving out the hold-out dataset, generating $\underline{v}$ as a sample of size 97 from $N\left(0,1\right)$ and setting $\underline{\mu} = {\underline{x}} -\underline{v}$, the model (\ref{E5}) could be rewritten as
\begin{eqnarray}\label{E6}
Y\left( {{s_i}} \right) = {\beta _0} + {\beta _1}{\mu \left( {{s_i}} \right)} + \sigma \varepsilon \left( {{s_i}} \right) + \sigma \omega {\rho \left( {{s_i}} \right)}+ {\beta_1}\sigma \tau {v \left( {{s_i}} \right)},
\end{eqnarray}
where $\tau=\sqrt{0.1}$ has been adopted. It must be mentioned that, in the following analysis, we assume that we have no information about ${\underline{x}}$ and $\underline{v}$ and just observed $\underline{\mu}$. The prior distributions are as follows: $\bbeta \sim {N_2}\left( {\underline{0},10{I_2}} \right)$, ${\sigma ^{ - 2}} \sim IG\left( {1.1,0.11} \right)$, ${\omega ^2} \sim  GIG\left( {0.001,0.05,2} \right)$, ${\tau ^2} \sim  GIG\left( {0,0.09,2} \right)$ and $\theta \sim Exp\left( 1 \right)$. Following consisting of comparing the performance of the measurement error model (MEM, hereafter) with naive model (NM, hereafter)
\begin{eqnarray}\label{E7}
Y\left( {{s_i}} \right) = {\beta _0} + {\beta _1}{\mu \left( {{s_i}} \right)} + \sigma \varepsilon \left( {{s_i}} \right) + \sigma \omega {\rho \left( {{s_i}} \right)}, 
\end{eqnarray}
which ignore the measurement error.

To implement computer programming, we need to set some initial (basic) values, so, for the two models (\ref{E6}) and (\ref{E7}) we set: $\beta_0=1.5$, $\beta_1=3$, $\sigma^2=2.8$, $\omega^2=3$,  $\tau^2=0.1$, $\theta=5$ and the initial values for the latent variables $\varepsilon$ and $V$ are two samples of size 97 which have been generated from ${\varepsilon _i}\stackrel{iid}{\sim} N\left( {0,0.31} \right)$ and ${v _i} \stackrel{iid}{\sim} N\left( {0,0.31} \right)$, respectively.

\begin{figure}
\centering
\includegraphics[width=0.95\textwidth, height=0.4\textwidth]{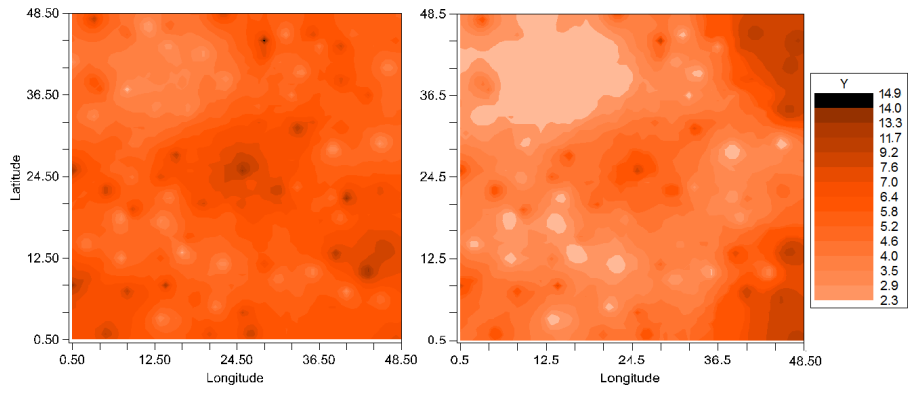}
%\vspace{-0.4cm}
\caption{Map of predicted values under MEM (left) and NM (right).}
\label{f4}
\end{figure}

\begin{figure}
\centering
\includegraphics[width=0.95\textwidth, height=0.4\textwidth]{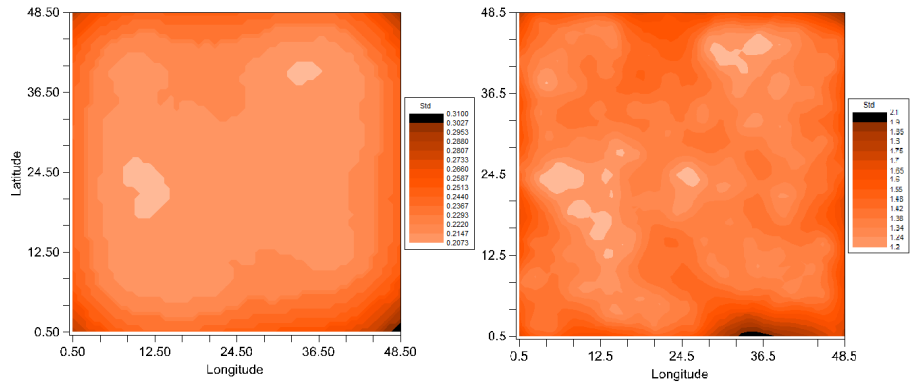}
\caption{Map of standard deviations of prediction under MEM (left) and NM (right).}
\label{f5}
\end{figure}

In the sensitivity analysis, we change the prior of one parameter at a time, say of parameter $\zeta_i$, and monitor the posterior distributions, focusing in particular on that of $\zeta_i$. As a simple measure of how the resulting posterior has changed, we compute the absolute value of the induced change in the marginal posterior mean of $\zeta_i$ and divide it by the standard deviation computed under the benchmark prior. We call this the ``relative change". If $\zeta_i$ is a vector, then we calculate this relative change for each element and report the maximum relative change (MRE) across elements. Table \ref{t2} lists the various values used for the alternative prior hyperparameters, as well as the maximum (across priors) relative change recorded for each parameter using the two considered models.
\begin{center}
[Table \ref{t2} about here.]
\end{center}
Results are based on $50,000$ draw from an MCMC chain after a burning period of $25,000$ iterations in which the taken lag value was $10$ to avoid the correlation problem in the generated chains. Convergence of the MCMC was verified through the Gelman and Rubin convergence diagnostics \cite{gelman1992inference} using the R package \textit{coda}. the Gelman and Rubin convergence diagnostics indicated convergence for each parameter and for a sample of the elements of the latent process. According to Table \ref{t2}, the prior on the model parameters  does not appear to play a very critical role based on measurement error model comparing with nive model. The results of models fitting are given in Table \ref{t3}. According to this table, it is found that the value of DIC criterion for the measurement error model is smaller than nive model and it shows that the proposed model has a better performance. In interpreting this table we could also argue that since the naive model is forced to modeling the variance of response just in terms of two elements $\sigma^2$ and $\omega^2$ while the measurement error model has a wider choice based on $\sigma^2$, $\omega^2$ and $\tau^2$, it is clear that the naive model faces with overestimation.
\begin{center}
[Table \ref{t3} about here.]
\end{center}
Furthermore, we study the sensitivity of two considered models to the choice of their initial values. Results which have been presented in Table \ref{t4}, shows the most stability of the results based on measurement error model. According to this table, by far the largest influence of initial value changes occurs for $\tau^2$, of course, this was not an unexpected problem from the beginning. Recall that $\tau^2$ is the parameter of the latent variable $V$ and so the model is powerless in the interpretation of this parameter comparing with the others.
\begin{center}
[Table \ref{t4} about here.]
\end{center}
Since the proposed model introduces the extra parameter $\tau^2$ beyond the parameterization of the nive model,  it is natural to examine to what extent information on this parameter can be recovered from data. Now, in order to address parameters identifiability, we focus on the  parameters $\sigma^2$, $\omega^2$ and $\tau^2$, because we would expect inference to be most challenging for these parameters and we generate three data sets with different values for them. Then, the estimates of the model parameters are obtained. Table \ref{t4a} displays the results.
\begin{center}
[Table \ref{t4a} about here.]
\end{center}
Finally, the contour map corresponding to the predictive mean and variance under the measurement error and naive models are shown in Figure \ref{f4} and Figure \ref{f5}, respectively.
\begin{center}
[Figure \ref{f4} about here.]
\end{center}
\begin{center}
[Figure \ref{f5} about here.]
\end{center}

\section{Piemonte dataset}\label{s4}
\begin{figure}
\centering
\includegraphics[width=\textwidth, height=0.7\textwidth]{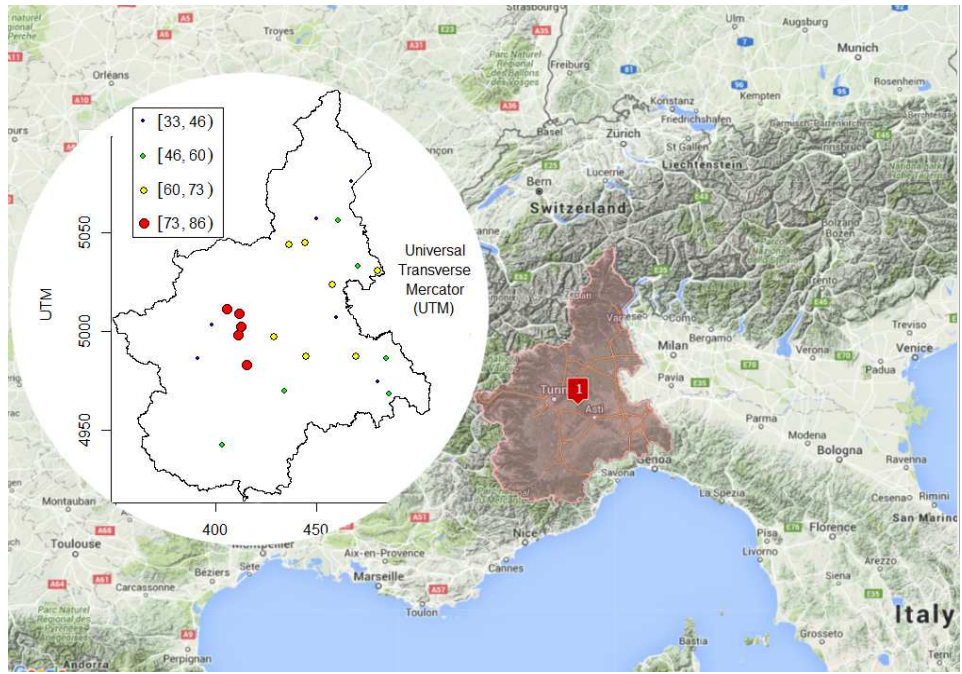}
\caption{The ${\rm PM}_{10}$ ($\mu g/{m^3}$) concentration data shown in a map of the Piemonte region. Filled circles are 24 registration sites of the ${\rm PM}_{10}$ ($\mu g/{m^3}$) concentration. Size of circles indicate the proportion of the day with the ${\rm PM}_{10}$ concentration.}
\label{f6}
\end{figure}

This section consists of an illustrative application of our methodology to investigate the spatial variations of particulate matter pollution, which is a serious concern in Piemonte north-western Po Valley, Italy. Since this region surrounded on three sides by the Alps, we usually observe lower ${\rm PM}_{10}$ concentration near the Alps, whereas higher pollution levels are detected in plains closer to urban areas. We focus on ${\rm PM}_{10}$ data come from a monitoring network composed of 24 stations in one day in November 2005. Figure \ref{f6} shows a schematic description of the region, locations of the stations and the ${\rm PM}_{10}$ concentration. The data contains ${\rm PM}_{10}$ concentration (in $\mu g/{m^3}$), spatial coordinates in terms of UTM (in $km$) and some covariates: (\textbf{I}) maximum mixing layer height (MLH, in $m$), (\textbf{II}) mean wind speed (WS, in $m/s$), (\textbf{III}) mean temperature (TEMP, in $K$), (\textbf{IV}) emission rates of primary aerosols (EMI, in $g/s$) and (\textbf{V}) altitude (A, in $m$). 
\begin{center}
[Figure \ref{f6} about here.]
\end{center}
(\textbf{I}) Mixing layer height  is an important parameter for understanding the transport process in the troposphere, air pollution, weather and climate change. Many methods have been proposed to determine MLH by identifying the turning point of the radiosonde profile. However, these methods are associated with the measurement error of humidity instruments and some other factors \citep{WangWang}.\\
(\textbf{II}) Rotation anemometers such as cup anemometers and rotation vane anemometers are apparently the most usual instruments for the measurement of the wind speed. A professional, well calibrated anemometer has at least a measurement error around $1\%$ for about 700-900 USD. Different types of errors in wind speed measurements by rotation anemometers were analysed by \cite{MacCready}.\\
(\textbf{III}) In temperature measurement, to obtain accuracies better than $0.2^\circ C$ great care is needed. Errors occur due to the presence of temperature gradients, drafts, sensor nonlinearities, poor thermal contact, calibration drifts, radiant energy and sensor self heating  \citep{Nakamura}.\\
(\textbf{IV}) Measuring the emission rate of an aerosol also faces with measurement error which studied by \cite{Bemeretal}.\\
(\textbf{V})Finally, the altitude as a GPS information faces with measurement error as well, which discussed in previously mentioned study \cite{Militino}.

\begin{figure}
        \centering
        \begin{subfigure}[t]{0.49\textwidth}
                \centering
                \includegraphics[width=0.6\textwidth, height=0.6\textwidth]{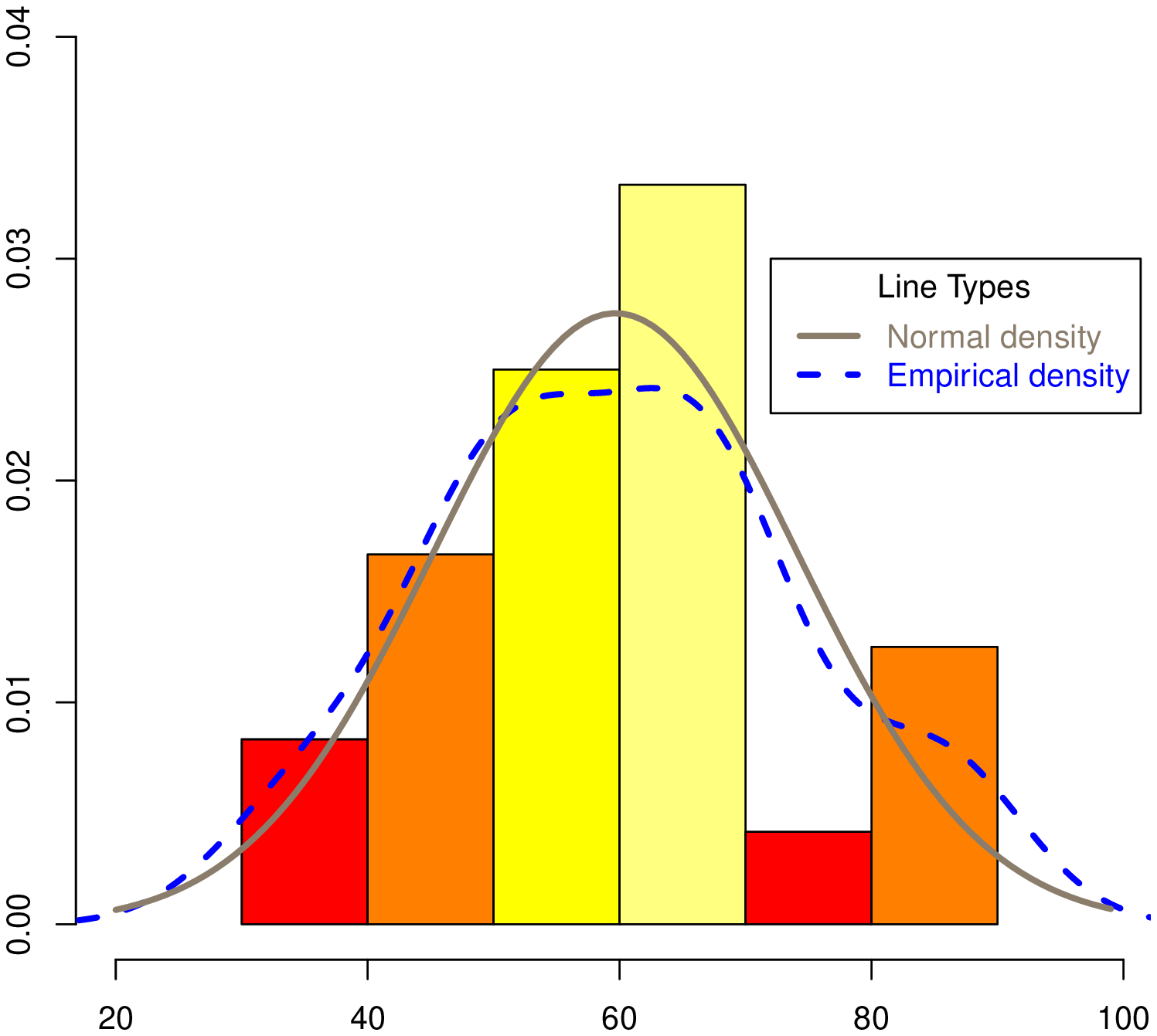}
                \caption{}
        \end{subfigure}%
        ~ %add desired spacing between images, e. g. ~, \quad, \qquad etc. 
        \begin{subfigure}[t]{0.48\textwidth}
                \centering
                \includegraphics[width=0.55\textwidth, height=0.55\textwidth]{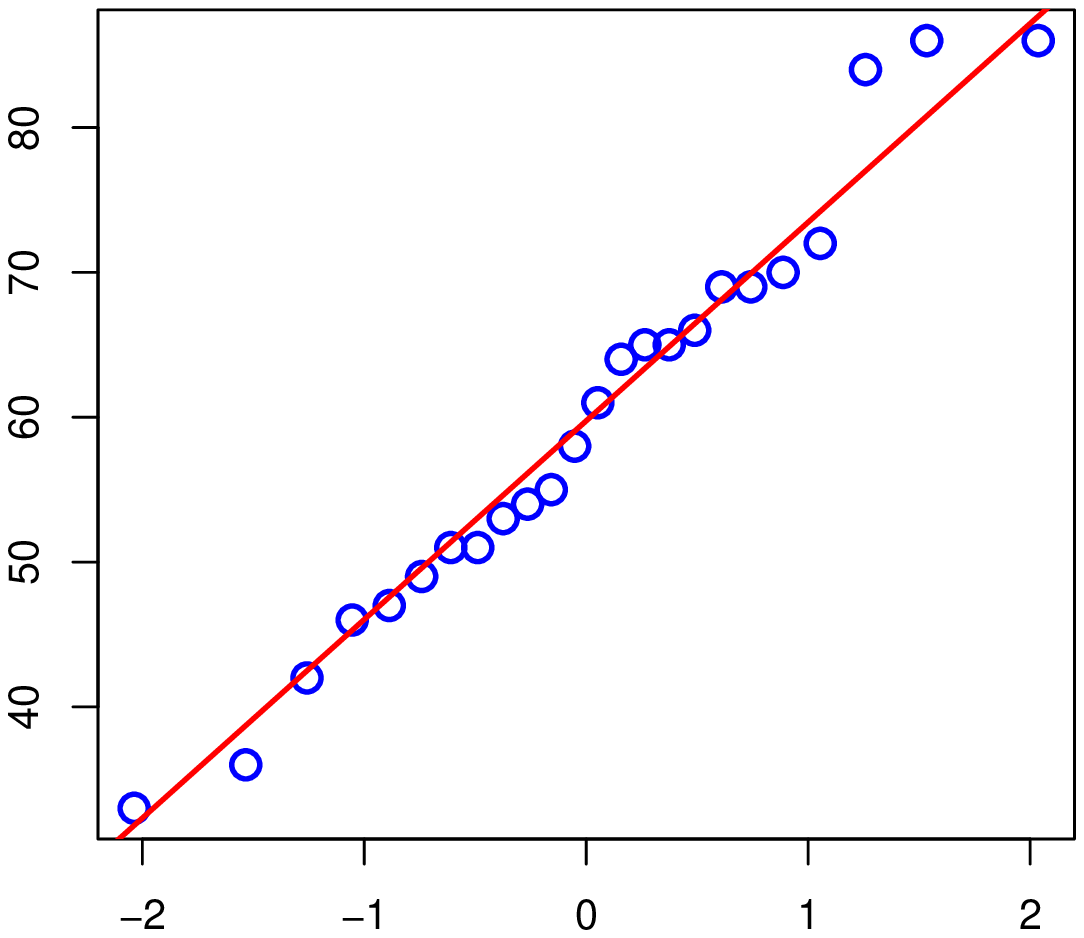}
                \caption{}
        \end{subfigure}
        \caption{(a) Histogram, (b) Q-Q plot.}\label{f7}
\end{figure}

\begin{figure}
        \centering
        \begin{subfigure}[t]{0.47\textwidth}
                \centering
                \includegraphics[width=0.86\textwidth, height=0.86\textwidth]{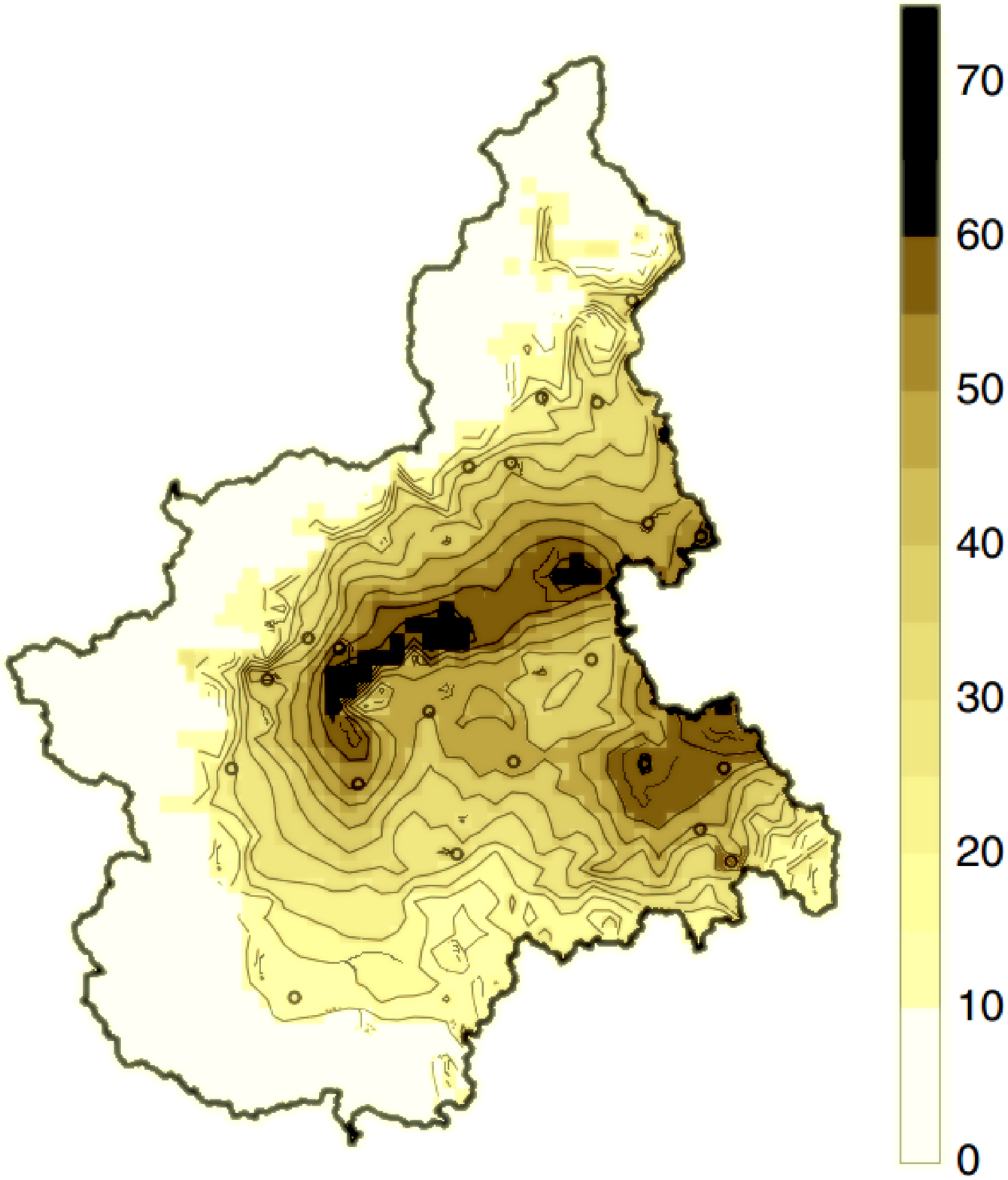}
                \caption{}
        \end{subfigure}%
        ~ %add desired spacing between images, e. g. ~, \quad, \qquad etc. 
        \begin{subfigure}[t]{0.52\textwidth}
                \centering
                \includegraphics[width=0.8\textwidth, height=0.8\textwidth]{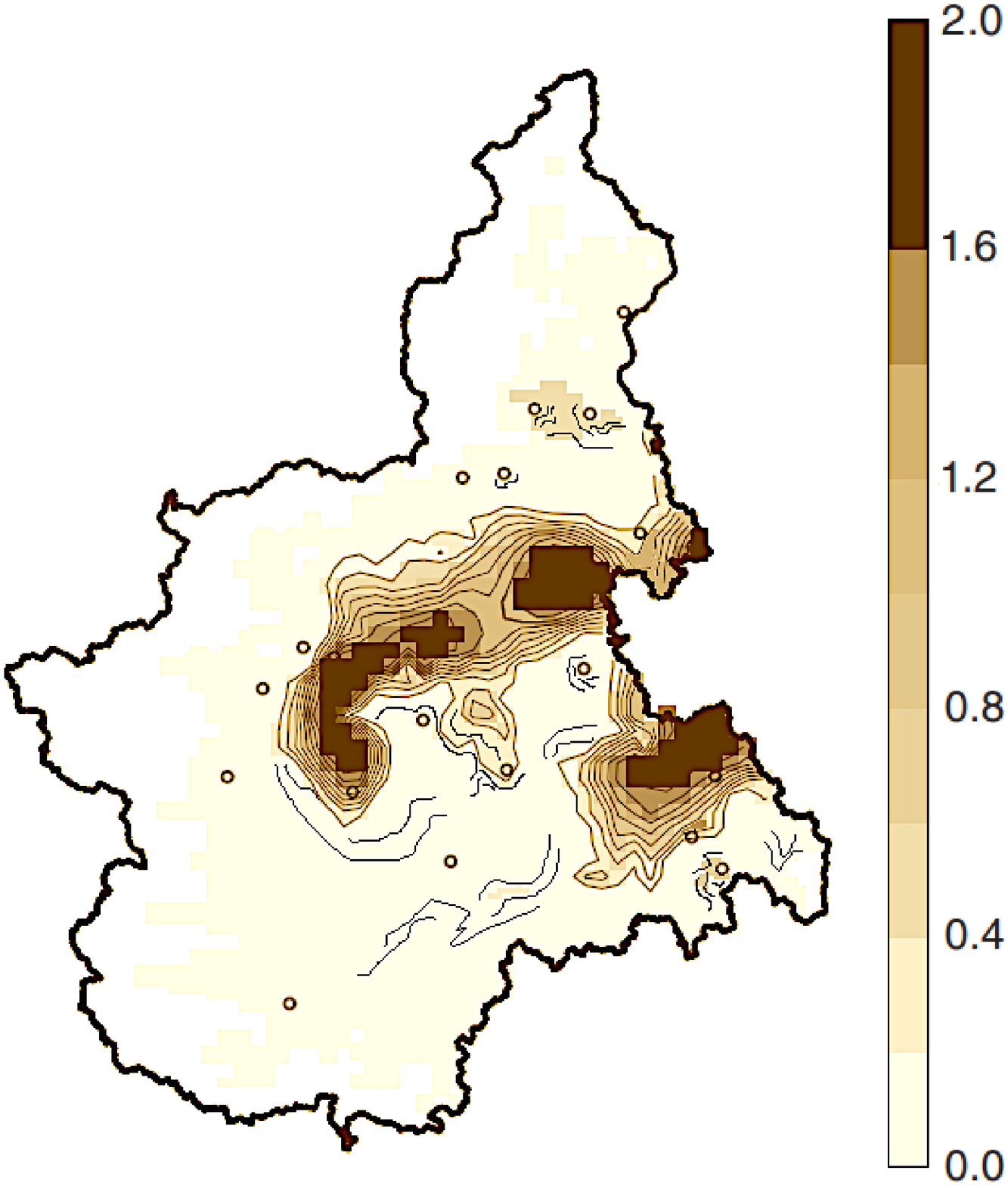}
                \caption{}
        \end{subfigure}
        \caption{(a) Map of predicted ${\rm PM}_{10}$ concentration,  (b) Uncertainty measure (standard deviations of prediction) of ${\rm PM}_{10}$ concentration.}\label{f9}
\end{figure}

To our knowledge, these aspects not only justify our methodology, but they are also a relevant issue for statistical applications. As mentioned in Section \ref{s1}, the main aim is to provide the reasonable prediction map based on information contained in the data. For this purpose, we view the data as a partial realization from a random field. Our initial exploratory analysis of the ${\rm PM}_{10}$ data did not find a serious violation of the normality assumption (see Figure \ref{f7} and remember that the sample size is small,) and so we advocate the use of Gaussian model (because of computational facilities).
\begin{center}
[Figure \ref{f7} about here.]
\end{center}
The prior for the model parameters is as described in Section \ref{s3} with ${c_1}=100$, ${c_2}=1.5$, ${c_3}=0.15$, ${c_4}=0.04$, ${c_5}=1.3$, ${c_6}=0.08$, ${c_7}=1.7$, ${c_8}=1.2$ and ${\gamma}=0.001$, where the prior for parameters is centered at a value close to the estimated value from the exact measurements. Table \ref{t5} reports all the posterior summaries.
\begin{center}
[Table \ref{t5} about here.]
\end{center}
Finally, the contour map corresponding to the predictive mean is shown in Figure \ref{f9}. According to this figure, the predictions are highest in plains closer to urban areas. The reason is low temperatures near the ground are often related to strong thermal inversion, one of the atmospheric features responsible for heavy pollution events in urban area. Besides, it is known that low temperatures cause an increase in particulate emissions from vehicle traffic sources.
\begin{center}
[Figure \ref{f9} about here.]
\end{center}

\section{Conclusion}\label{s5}
In this paper, we have proposed spatial linear measurement error model to account for covariate measurement error and spatial correlation in spatial data. For this purpose, we adopted the Bayesian approach and utilized the Markov chain Monte Carlo algorithms and data augmentations to carry out calculations. Using simulated data, meaningful inference on parameters, prediction, sensitivity to the priors and identifiability was concluded.

Since all covariates in ${\rm PM}_{10}$ concentration faced with instrumental measurement error, this problem indicates that our methodology is well-suited to analyze these data. Obviously, the conclusions and the final suggestion in our case study do not constitute the general answer to the question about the best choice of computational methods. Recently, some studies have been performed to find replacing methods and it has led to new methods. For example, the Variational Bayes’ method is one of the most ones. Although this method requires more complex theoretic calculations, it could increase the speed of calculations.

\appendix
\section*{Appendix: The conditional distributions}\label{App1}
Below is the full conditional distributions of all unobservable quantities to draw samples from $P\left( {{\eta},\bvarepsilon ,V \left| \underline{y} \right.} \right)$ in the Gibbs sampler framework. In what follows, we use the notation ${{\eta _{ - \phi }}}$ to show the vector $\eta$ without $\phi$. Furthermore, we consider the vectorized version of the matrices $\mu {\left( s \right)_{n \times p}}$ and $V {\left( s \right)_{n \times p}}$ as $\mu {\left( s \right)_{n \times p}} = \left( {{\mu _1},{\mu _2}, \ldots, {\mu _p}} \right)$ and $V{\left( s \right)_{n \times p}} = \left( {{v_1},{v_2}, \ldots ,{v_p}} \right)$, respectively. Regardless of the details, the full conditional distributions are as follows:

\begin{itemize}
\item[$\bullet$] \textbf{Full conditional distribution of latent variable $\bvarepsilon$}
\begin{eqnarray*}
P\left( {\bvarepsilon \left| {v,\eta ,\underline{y}} \right.} \right) &\propto &P\left( {\underline{Y}\left| {\bvarepsilon ,v,\eta } \right.} \right)P\left( {\bvarepsilon \left| \eta  \right.} \right)\\
&\propto &\exp \left\{ { - \frac{1}{2}{{\left( {\bvarepsilon  - {A_1^{ - 1}}\textbf{z}^* } \right)}^\prime }{A_1}\left( {\bvarepsilon  - {A_1^{ - 1}}\textbf{z}^* } \right)} \right\},
\end{eqnarray*}
where $A_1 = \frac{{{1}}}{{{\omega ^2} }}{I_n} + C_\theta ^{ - 1}$ and $\textbf{z}^*$ is a $1\times n$ vector with elements ${z_i} = \frac{{{y_i} - {\mu _i^\prime }\bbeta  - {{\sigma}{\tau}}{v_i^\prime }\bbeta }}{{\omega^2}\sigma }$. So, $\bvarepsilon \left| {v,\eta ,} \right.\underline{y} \sim {N_n}\left( {{A_1^{ - 1}}\textbf{z}^* ,{A_1^{ - 1}}} \right)$.
%%%%%%%%%%%%%%%%%%%%%%%%%%%%%%%%%%%%
%%%%%%%%%%%%%%%%%%%%%%%%%%%%%%%%%%%%
%%%%%%%%%%%%%%%%%%%%%%%%%%%%%%%%%%%%
\item[$\bullet$] \textbf{Full conditional distribution of latent variable $V$}\\
If ${r_i} = \frac{{{y_i} - {\mu _i^\prime }\bbeta  - \sigma {\bvarepsilon _i}}}{{{\sigma}{\tau }}}$ and $\textbf{r} ' = \left( {{r_1},{r_2}, \ldots ,{r_n}} \right)$, we have
\begin{eqnarray*}
P\left( {V\left| {\bvarepsilon ,\eta ,\underline{y}} \right.} \right) &\propto &P\left( {\underline{Y}\left| {\bvarepsilon ,v,\eta } \right.} \right)P\left( {V\left| \eta  \right.} \right)\\
&\propto &\exp \left\{ { - \frac{1}{2}\left[ {\frac{{\tau^2}}{{{\omega ^2} }}\left[ {{{\left( {V\bbeta } \right)}^\prime }\left( {V\bbeta } \right) - 2{{\left( {V\bbeta } \right)}^\prime }\textbf{r} } \right] + V'V} \right]} \right\},
\end{eqnarray*}
which has not a standard form, so we determine $P\left( {V\bbeta\left| {\bvarepsilon ,\eta ,\underline{y}} \right.} \right)$ and then using customary approach to solve under-determined systems of linear equations \citep{Donoho, Donoho2} we can draw from $P\left( {V\left| {\bvarepsilon ,\eta ,\underline{y}} \right.} \right)$. Thus,
\begin{eqnarray*}
P\left( {V\bbeta\left| {\bvarepsilon ,\eta ,\underline{y}} \right.} \right) &\propto &P\left( {\underline{Y}\left| {\bvarepsilon ,v,\eta } \right.} \right)P\left( {V\bbeta\left| \eta  \right.} \right)\\
&\propto &\exp \left\{ { - \frac{1}{2}{{\left( {V\bbeta  - A_2^{ - 1}\textbf{r}^* } \right)}^\prime }{A_2}\left( {V\bbeta  - A_2^{ - 1}\textbf{r}^*} \right)} \right\},
\end{eqnarray*}
where $A_2 = {\left( {\frac{{\tau^2}}{{{\omega ^2} }} + 1} \right){I_n}}$ and $\textbf{r}^*=\frac{\tau^2}{\omega^2}\textbf{r}$. Hence, 
\begin{eqnarray*}
V\bbeta \left| {\bvarepsilon ,\eta ,} \right.\underline{y}\sim{N_n}\left( {A_2^{ - 1}\textbf{r}^* ,A_2^{ - 1}} \right).
\end{eqnarray*}
%%%%%%%%%%%%%%%%%%%%%%%%%%%%%%%%%%%%
%%%%%%%%%%%%%%%%%%%%%%%%%%%%%%%%%%%%
%%%%%%%%%%%%%%%%%%%%%%%%%%%%%%%%%%%%
\item[$\bullet$] \textbf{Full conditional distribution of parameter $\bbeta$}\\
By setting ${t_i} = {y_i} - \sigma {\varepsilon _i}$, $\textbf{t} ' = \left( {{t_1},{t_2}, \ldots {t_n}} \right)$, ${t_i^*}' = {\mu _i^\prime } + {{\sigma }{\tau }}{v_i^\prime }$ and ${\textbf{T} ^*}'  = \left( {{t_1^*}',{t_2^*}', \ldots {t_n^*}'} \right)$, we have
\begin{eqnarray*}
\pi \left( {\bbeta \left| {\bvarepsilon ,v,{\eta _{ - \bbeta }},\underline{y}} \right.} \right)  &\propto&  P\left( {\underline{Y}\left| {\bvarepsilon ,v,\eta } \right.} \right)\pi \left( \bbeta  \right)\\
&\propto &\exp \left\{ { - \frac{1}{2}\left[ {\bbeta '\left( {\frac{{{\textbf{T}^*}^\prime {\textbf{T}^*}}}{{{\sigma ^2}{\omega ^2}}} + \frac{{{I_p}}}{{{c_1}}}} \right)\bbeta  - 2\bbeta '\frac{{{\textbf{T}^*}^\prime\textbf{t}}}{{{\sigma ^2}{\omega ^2}}}} \right]} \right\}\\
&\propto &\exp \left\{ { - \frac{1}{2}{{\left( {\bbeta  - A_3^{ - 1}F} \right)}^\prime }{A_3}\left( {\bbeta  - A_3^{ - 1}F} \right)} \right\}.
\end{eqnarray*}
Therefore, $\bbeta \left| {\bvarepsilon ,v,{\eta _{ - \bbeta }},\underline{y}} \right. \sim {N_p}\left( {A_3^{ - 1}F,A_3^{ - 1}} \right)$ with 
\begin{eqnarray*}
{A_3} = \left( {\frac{1}{{{\sigma ^2}{\omega ^2}}}{\textbf{T}^*}^\prime {\textbf{T}^*} + \frac{1}{{{c_1}}}{I_p}} \right)
\end{eqnarray*}
and
${F = \frac{1}{{{\sigma ^2}{\omega ^2}}}{\textbf{T}^*}^\prime \textbf{t}}$.
%%%%%%%%%%%%%%%%%%%%%%%%%%%%%%%%%%%%
%%%%%%%%%%%%%%%%%%%%%%%%%%%%%%%%%%%%
%%%%%%%%%%%%%%%%%%%%%%%%%%%%%%%%%%%%
\item[$\bullet$] \textbf{Full conditional distribution of parameter $\sigma^2$}
\begin{eqnarray*}
\pi \left( {{\sigma ^2}\left| {\bvarepsilon ,v,{\eta _{ - {\sigma ^2}}},\underline{y}} \right.} \right) &\propto & P\left( {\underline{Y}\left| {\bvarepsilon ,v,\eta } \right.} \right)\pi \left( {{\sigma ^2}} \right)\\
&\propto&{\left( {\frac{1}{{{\sigma ^2}}}} \right)^{\frac{n}{2} + {c_2} + 1}}\exp \left\{ { - \frac{1}{{2{\omega ^2}}}\sum\limits_{i = 1}^n {{{\left( {{q_i^*} - \frac{q_i}{\sigma}} \right)}^2} - \frac{{{c_3}}}{{{\sigma ^2}}}} } \right\},
\end{eqnarray*}
in which ${q_i} = {y_i} - {\mu _i}^\prime \bbeta$ and $q_i^* = {\varepsilon _i} + \tau {v'_i}\bbeta $.
This full conditional distribution is of nonstandard form, so a Metropolis-Hastings step or sampling importance resampling method would be used.
%%%%%%%%%%%%%%%%%%%%%%%%%%%%%%%%%%%%
%%%%%%%%%%%%%%%%%%%%%%%%%%%%%%%%%%%%
%%%%%%%%%%%%%%%%%%%%%%%%%%%%%%%%%%%%
\item[$\bullet$] \textbf{Full conditional distribution of parameter $\omega^2$}
\begin{eqnarray*}
\pi \left( {{\omega ^2}\left| {\bvarepsilon ,v,{\eta _{ - {\omega ^2}}},\underline{y}} \right.} \right) &\propto & P\left( {\underline{Y}\left| {\bvarepsilon ,v,\eta } \right.} \right)\pi \left( {{\omega ^2}} \right)\\
&= &{\left( {{\omega ^2}} \right)^{\gamma  - \frac{n}{2} - 1}}\exp \left\{ { - \frac{1}{2}\left[ {\frac{{{d^*}}}{{{\omega ^2}}} + c_5^2{\omega ^2}} \right]} \right\},
\end{eqnarray*}
where ${d_i} = {y_i}-{\mu '_i}\bbeta -\sigma {\varepsilon _i} - {{\sigma }{\tau }}{v'_i}\bbeta $ and ${d^*} ={{c_4^2}+\frac{1}{\sigma^2} \sum\limits_{i = 1}^n {d_i^2}} $. Hence, we have ${\omega ^2}\left| {\bvarepsilon ,v,{\eta _{ - {\omega ^2}}},\underline{y} \sim GIG\left( {\gamma  - \frac{n}{2},\sqrt{d^*},{c_5}} \right)} \right.$.
\end{itemize}
In the three last items, the full conditional distribution of parameter $\tau^2$, $\theta_1$ and $\theta_2$, are of nonstandard form, so a Metropolis-Hastings step or sampling importance resampling method would be used.
\begin{itemize}
\item[$\bullet$] \textbf{Full conditional distribution of parameter $\tau^2$}\\
By setting $r_i^* = \frac{1}{\sigma}\left( {y_i}-{\mu '_i}\bbeta -\sigma {\varepsilon _i}\right) $, we can write
\begin{eqnarray*}
\pi \left( {{\tau ^2}\left| {\bvarepsilon ,v,{\eta _{ - {\tau ^2}}},\underline{y}} \right.} \right) \hspace{-3mm}&\propto &\hspace{-3mm} P\left( {\underline{Y}\left| {\bvarepsilon ,v,\eta } \right.} \right)\pi \left( {{\tau ^2}} \right)\\
&\propto &\hspace{-3mm} \frac{1}{{{\tau ^2}}}\exp \left\{ { - \frac{1}{2}\left[ {\frac{{c_6^2}}{{{\tau ^2}}} + c_7^2{\tau ^2} + \frac{{{\sigma ^2}}}{{{\sigma ^2}{\omega ^2}}}\sum\limits_{i = 1}^n {{{\left( {r_i^* - \tau v_i^\prime \bbeta } \right)}^2}} } \right]} \right\}.
\end{eqnarray*}
%%%%%%%%%%%%%%%%%%%%%%%%%%%%%%%%%%%%
%%%%%%%%%%%%%%%%%%%%%%%%%%%%%%%%%%%%
%%%%%%%%%%%%%%%%%%%%%%%%%%%%%%%%%%%%
\item[$\bullet$] \textbf{Full conditional distribution of parameter $\theta_1$}
\begin{eqnarray*}
\pi \left( {{\theta _1}\left| {\bvarepsilon ,v,{\eta _{ - {\theta _1}}},\underline{y}} \right.} \right) &\propto &P\left( {\bvarepsilon \left| {v,{\eta _{ - {\theta _1}}}} \right.} \right)\pi \left( {{\theta _1}} \right)\\
&\propto &{\left| {{C_{\left( {{\theta _1},{\theta _2}} \right)}}} \right|^{ - \frac{1}{2}}}\exp \left\{ { - \frac{1}{2}{\bvarepsilon ^\prime }C_{\left( {{\theta _1},{\theta _2}} \right)}^{ - 1}\bvarepsilon  - \frac{{{c_8}}}{{med\left( d \right)}}{\theta _1}} \right\}.
\end{eqnarray*}
%%%%%%%%%%%%%%%%%%%%%%%%%%%%%%%%%%%%
%%%%%%%%%%%%%%%%%%%%%%%%%%%%%%%%%%%%
%%%%%%%%%%%%%%%%%%%%%%%%%%%%%%%%%%%%
\item[$\bullet$] \textbf{Full conditional distribution of parameter $\theta_2$}
\begin{eqnarray*}
\pi \left( {{\theta _2}\left| {\bvarepsilon ,v,{\eta _{ - {\theta _2}}},\underline{y}} \right.} \right) &\propto &P\left( {\bvarepsilon \left| {v,{\eta _{ - {\theta _2}}}} \right.} \right)\pi \left( {{\theta _2}} \right)\\
&\propto &{\left| {{C_{\left({\theta_1},{\theta_2}\right) }}} \right|^{ - \frac{1}{2}}}\exp \left\{ { - \frac{1}{2}\bvarepsilon 'C_{\left({\theta_1},{\theta_2}\right)} ^{ - 1}\bvarepsilon -{  {{{c_9}}}{\theta _2}} } \right\}.
\end{eqnarray*}
\end{itemize}

%%%%%%%%%%%%%%%%%%%%%%%%%%%
%{\small
%\bibliographystyle{cJAS}
%\bibliography{cJASguide}}

%\newpage
%\begin{figure}
%\centering
%\includegraphics[width=0.65\textwidth, height=0.5\textwidth]{Figure1.eps}
%\caption{${\rm PM}_{10}$ is about ${1/7}^{th}$ the diameter of a human hair and ${\rm PM}_{2.5}$ is about ${1/28}^{th}$ the diameter of a human hair.}
%\label{f1}
%\end{figure}
%

\clearpage 
\newpage
\begin{table}
\centering
\caption{The coordinates of 11 incorporated locations.}
{\begin{tabular}{| c | c c || c | c c |}
\hline%\toprule
{\footnotesize Num.} & Longitude & Latitude & {\footnotesize Num.} & Longitude & Latitude \\
\hline%\cmidrule(lr){1-6}
1&$14.143578$ & $8.449528$&7&$22.677188$ & $38.815286$\\
2&$13.610791$ & $17.782726$&8&$33.783750$ & $39.866913$\\
3&$9.004231$   & $24.223948$&9&$33.151787$ & $23.054762$\\
4&$8.507509$   & $37.369297$&10&$43.535390$ & $36.435227$\\
5&$18.034563$ & $27.378832$&11&$38.923097$ & $13.050401$\\
6&$24.829648$ & $20.148889$&&&\\
\hline%\bottomrule
\end{tabular}}\label{t1}
\end{table}

\clearpage
\newpage

\clearpage
\newpage
\begin{table}
\centering 
\caption{Prior sensitivity analysis: setup of the experiment and  relative change  based on both measurement error and naive models, (MEM) and (NM), respectively.}
{\begin{tabular}{c c c c c c c c c c} 
\toprule 
&&&&\multicolumn{2}{c}{\footnotesize{Relative change }}\\
\cmidrule{5-6} 
			\footnotesize{$\zeta_i$}
			& \footnotesize{Hyperparameter}
			& \footnotesize{Benchmark}
			& \footnotesize{Alternative values}
			&\footnotesize{NM}
			&\footnotesize{MEM }\\ 
\cmidrule{1-6} 
${\beta_0}$ & $c_1$ & $10$ & $9$ & $0.2097$ &$0.1989$
\\ 
${\beta_1}$ & $c_1$ & $10$ & $13$ & $0.2462$ & $0.2144$
\\
${\sigma^2}$ & $\left({c_2},{c_3}\right)$ & $\left(1.1,0.11\right)$ & $\left(0.09,0.19\right)$ & $0.798$
& $0.6955$
\\
${\omega^2}$ & $\left(\gamma,{c_4},{c_5}\right)$ & $\left({10^{-3}},{0.05},{2}\right)$ & $\left({10^{-3}},{0.03},{1.7}\right)$ & $0.3564$ & $0.3613$
\\
$\theta$ & $c_8$ & $1$ & $1.28$ & $0.3221$ & $0.303$
\\
$\tau^2$ & $\left({c_6},{c_7}\right)$ & $\left({0.09},{2}\right)$ & $\left(0.05 ,1.46\right)$ & -
& $0.728$
\\
\bottomrule
\end{tabular}}
\label{t2}
\end{table}

\clearpage
\newpage
\begin{table}
\centering
\caption{The estimated value (EVal) and the estimated variance ({\textcolor{blue}{EVar}}) of parameters with the absolute value of the difference ({\textcolor{red}{AVD}}) between true value (TV) and Eval  based on both measurement error and naive models, (MEM) and (NM), respectively.}
{\begin{tabular}{c c c c c c c c c}
\toprule 
 &&
\multicolumn{3}{c}{\footnotesize{NM }}&
\multicolumn{3}{c}{\footnotesize{MEM}}\\
\cmidrule(rl){3-5} \cmidrule(rl){6-8} 
		\footnotesize{Parameter}
		&\footnotesize{TV}
		&\footnotesize{EVal}
		&\footnotesize{\textcolor{red}{AVD}}
		&\footnotesize{\textcolor{blue}{EVar}}
		&\footnotesize{EVal}
		&\footnotesize{\textcolor{red}{AVD}}
		&\footnotesize{\textcolor{blue}{EVar}}\\ 
\cmidrule{1-8} 
${\beta_0}$& $0.5$&$-0.6571$
&{\textcolor{red}{$1.1571$}}&{\textcolor{blue}{$1.087$}}&$0.2119$&{\textcolor{red}{$0.2881$}}&{\textcolor{blue}{$0.96$}}
\\ 
${\beta_1}$& $2$&$2.9751$
&{\textcolor{red}{$0.9751$}}&{\textcolor{blue}{$0.955$}}&$2.1196$&{\textcolor{red}{$0.1196$}}&{\textcolor{blue}{$0.971$}}
\\
${\sigma^2}$& $1$&$4.0176$
&{\textcolor{red}{$3.0176$}}&{\textcolor{blue}{$2.862$}}&$1.5047$&{\textcolor{red}{$0.5047$}}&{\textcolor{blue}{$1.313$}}
\\
${\omega^2}$& $1.1$&$0.029$
&{\textcolor{red}{$1.071$}}&{\textcolor{blue}{$0.409$}}&$0.3788$&{\textcolor{red}{$0.7212$}}&{\textcolor{blue}{$1.025$}}
\\
$\theta$& $1.2$&$1.3721$
&{\textcolor{red}{$0.1721$}}&{\textcolor{blue}{$1.263$}}&$1.3719$&{\textcolor{red}{$0.1719$}}&{\textcolor{blue}{$1.532$}}
\\
$\tau^2$& $0.1$ &-
&{\textcolor{red}{-}}&{\textcolor{blue}{-}}&$0.2301$&{\textcolor{red}{$0.1301$}}
&{\textcolor{blue}{$0.0898$}}
\\
\cmidrule(rl){3-5} \cmidrule(rl){6-8} 
\multicolumn{2}{c}{\footnotesize{{\textcolor{blue}{DIC}}}}& \multicolumn{3}{c}{\footnotesize{{\textcolor{blue}{$402.13$}}}} & \multicolumn{3}{c}{\footnotesize{{\textcolor{blue}{$351.04$}}}}\\
\bottomrule
\end{tabular}}
\label{t3}
\end{table}

\clearpage
\newpage
\begin{table}
\centering 
\caption{Methodology sensitivity analysis: setup of the experiment and maximum relative change ({\textcolor{red}{MRE}}) based on both measurement error and naive models, (MEM) and (NM), respectively.}
{\begin{tabular}{c c c c c c c c c c} 
\toprule 
&&&&\multicolumn{2}{c}{\footnotesize{\textcolor{red}{{\textcolor{red}{MRE}}}}}\\
\cmidrule{5-6} 
			\footnotesize{Parameter}
			& \footnotesize{Real value }
			& \footnotesize{Initial value }
			& \footnotesize{Alternative values }
			&\footnotesize{NM }
			&\footnotesize{MEM  }\\ 
\cmidrule{1-6} 
${\beta_0}$&$0.5$& $1.5$&$0.9$ , $0.1$&{\textcolor{red}{$0.2201$}}&{\textcolor{red}{$0.2113$}}
\\ 
${\beta_1}$&$2$& $3$&$1.11$ , $0.87$&{\textcolor{red}{$0.1756$}}
&{\textcolor{red}{$0.0189$}}
\\
${\sigma^2}$&$1$& $2.8$&$1.5$ , $0.5$&{\textcolor{red}{$0.4105$}}
&{\textcolor{red}{$0.2643$}}
\\
${\omega^2}$&$1.1$& $3$&$2.1$ , $0.76$&{\textcolor{red}{$0.3423$}}
&{\textcolor{red}{$0.2518$}}
\\
$\theta$&$1.2$& $5$&$3.5$ , $1.6$&{\textcolor{red}{$0.3061$}}
&{\textcolor{red}{$0.2714$}}
\\
$\tau^2$&$0.1$& $0.1$ &$0.4$ , $0.05$&{\textcolor{red}{-}}
&{\textcolor{red}{$0.31$}}
\\
\bottomrule
\end{tabular}}
\label{t4}
\end{table}

\clearpage
\newpage
\begin{table}
\centering
\caption{Identifiability of the parameters $\sigma^2$, $\omega^2$ and $\tau ^2$.}
\begin{tabular}{c l c c c c}
\toprule
\multirow{2}{*}{\textcolor{blue}{\large $\sigma^2$}}
 &   True value  & $0.85$ & $1$ &$1.33$ \\
 & Estimated value & $1.091$ & $1.493$ &$1.2755$  \\
\hline
\multirow{2}{*}{\textcolor{blue}{\large $\omega ^2$}}
 &   True value & $0.74$ & $1.1$ &$1.28$ \\
 &  Estimated value & $0.9882$ & $0.6911$ & $1.1704$  \\
\hline
\multirow{2}{*}{\textcolor{blue}{\large $\tau ^2$}}
 &   True value & $0.08$ & $0.1$ &$0.19$ \\
 &  Estimated value & $0.0957$ & $0.3006$ & $0.2126$  \\
\bottomrule
\end{tabular}
\label{t4a}
\end{table}

\clearpage
\newpage

\clearpage
\newpage

\clearpage
\newpage

\clearpage
\newpage
\begin{table}
%\centering 
\caption{The estimated value (EVal) and the estimated variance (EVar) of parameters for
the Piemonte air pollution analysis.}
{\begin{tabular}{m{5em}  c c   c c c c c c c} 
\toprule 
			\footnotesize{\textcolor{blue}{Parameter}}
			& \footnotesize{EVal}
			& \footnotesize{EVar}
			&\footnotesize{\textcolor{blue}{Parameter}}
			& \footnotesize{EVal }
			& \footnotesize{EVar}\\ 
\cmidrule{1-6} 
{\textcolor{blue}{${\beta_0}$}} & $2.712$ & $1.015$ &{\textcolor{blue}{${\sigma^2}$}} & $2.051$ & $1.686$
\\ 
{\textcolor{blue}{${\beta_1}$}}{\textcolor{red}{(MLH)}} & $-0.033$ & $0.411$ &{\textcolor{blue}{${\omega^2}$}}  & $0.17$ & $0.871$
\\
{\textcolor{blue}{${\beta_2}$}}{\textcolor{red}{(WS)}}& $-0.026$ & $0.38$ & {\textcolor{blue}{$\theta$}} & $0.981$ & $0.604$
\\
{\textcolor{blue}{${\beta_3}$}}{\textcolor{red}{(TEMP)}} & $-0.197$ & $0.405$ & {\textcolor{blue}{$\tau^2$}} & $0.197$ & $0.893$
\\
{\textcolor{blue}{${\beta_4}$}}{\textcolor{red}{(EMI)}} & $0.119$ & $0.252$ & 
\\
{\textcolor{blue}{${\beta_5}$}}{\textcolor{red}{(A)}} & $-0.316$ & $0.109$ &  & &
\\
\bottomrule
\end{tabular}}
\label{t5}
\end{table}

\clearpage
\newpage

%\newpage
%\begin{figure}
%\centering
%\includegraphics[width=0.55\textwidth, height=0.55\textwidth]{Figure8.eps}
%\caption{Piemonte grid. Filled circles denote the 24 monitoring stations.}
%\label{f8}
%\end{figure}

\clearpage
\newpage

\end{document}